\documentclass[10pt,a4paper]{article}
\usepackage[latin1]{inputenc}
\usepackage{epsf}
\usepackage{amsmath}
\usepackage{amsfonts}
\usepackage{amssymb}
\usepackage{graphicx}
\usepackage{makeidx}
\begin{document}
\author{Subhankar Roy\footnote{Email: meetsubhankar @ gmail.com} and N. Nimai Singh\footnote{E-mail: nimai03 @ yahoo.com}\\ \\
Department of Physics, Gauhati University, Guwahati-781014, India }
\title{Expansion of $U_{PMNS}$ and Neutrino mass matrix $M_{\nu}$ in terms of $\sin\theta_{13}$ for Inverted Hierarchical case.}
\maketitle
\abstract
The recent observational data supports the deviation from
Tri-bimaximal (TBM) mixings. Different theoretical models suggest the
interdependency 
among the observational parameters involving the mixing angles. On
phenomenological ground
 we try to construct the PMNS matrix $U_{PMNS}$ with certain analytic
 structure satisfying 
 the unitary condition, in terms of a single observational parameter
 $\sin\theta_{13}$.
 We hypothesise the three neutrino masses, $m_{i}$ as functions of
 $\sin\theta_{13}$
 and then construct the neutrino mass matrix $M _{\nu}$. We assume the convergence of the model to TBM
 mixing when $\theta_{13}$ is taken zero. The mass matrix so far obtained 
can be employed for various applications including the estimation of matter-antimatter asymmetry of the universe.
\newpage
\section{Introduction}
Recent results published by Double Chooz[1], Daya Bay[2], RENO[3],
T2K[4] and MINOS[5] collaborations
 assure relatively large reactor angle ($\theta_{13}$). Also the
 recent global neutrino oscillation data analysis[ 6] insists on
 $\theta_{23}<\pi/4$. 
Tri-bimaximal Mixing[7] is associated with $\theta_{13}=0$, and
$\theta_{23}=\pi/4$. 
This  symmetry has a strong theoretical support because of its
relation with so called
 $\mu-\tau$ symmetry of neutrino mass matrix. $\mu-\tau$ symmetry
which is associated with $A_{4}$ discrete flavour symmetry
group[8-12].
 But in order to comply with the recent experimental results, some
 perturbations
 have to be introduced in this mixing pattern. An open question is
 whether 
the corrections[13,14] are needed or a new mixing scheme is to be introduced[15].

In the present literature[16,17] we find the dependency of the mixing
angles on one another.
 If this is true, then we are allowed to choose a single parameter
 capable of describing 
all the three mixing angles. We move a step ahead and express the
three masses 
under this parameter. This helps us to define a simplified neutrino
mass 
model with a single parameter only. 

Out of all the three observational parameters concerning the mixing
angles, $\sin\theta_{13}$ 
is the smallest one. So, we choose $\sin\theta_{13}$ as the guiding
parameter. 
We consider tri-bimaximal mixing pattern and $\mu-\tau $ symmetry as 
the first approximation. Hence the model is supposed to produce T.B.M 
mixing when we put $\sin\theta_{13}=0$. We try to keep the structure
of 
the three rotation matrices $U(\theta_{13})$, $U (\theta_{12})$ and
$U(\theta_{23})$ 
in analytical form so that they can satisfy the unitary condition 
$[U(\theta_{ij})]^{\dagger}U(\theta_{ij}) = I$. 
We start with the following ansatz,
\begin{eqnarray}
s_{13}&=&\epsilon ,\\
s_{12}&=&\frac{1}{\sqrt{3}}-\frac{\epsilon}{5},\\
s_{23}&=&\frac{1}{\sqrt{2}}-\frac{\epsilon}{2}.
\end{eqnarray}
where, $s_{ij}=\sin\theta_{ij}$, and then construct the PMNS mixing matrix and then the neutrino mass matrix in the usual way.
\section{Construction of the PMNS matrix}
We consider the charged lepton mass matrix to be diagonal. Hence we can choose $U_{PMNS}=U_{\nu}$. We propose the three rotation matrices as:
\begin{eqnarray}
U(\theta_{13})&=&\begin{pmatrix}
(1-\epsilon ^{2})^\frac{1}{2} & 0 & \epsilon  e^{-i\delta} \\ 
0 & 1 & 0 \\ 
-\epsilon  e^{i \delta} &0 & (1-\epsilon ^{2})^\frac{1}{2}
\end{pmatrix},\\
U(\theta_{23})&=&\begin{pmatrix}
1 & 0 & 0 \\ 
0 & (\frac{1}{2}+\frac{\epsilon}{\sqrt{2}}-\frac{\epsilon^{2}}{4})^\frac{1}{2} & \frac{\epsilon}{2}-\frac{1}{\sqrt{2}} \\
0 &\frac{1}{\sqrt{2}}-\frac{\epsilon}{2} &(\frac{1}{2}+\frac{\epsilon}{\sqrt{2}}-\frac{\epsilon^{2}}{4})^\frac{1}{2}
\end{pmatrix} \\
U(\theta_{12})&=&\begin{pmatrix}
(\frac{2}{3}+\frac{2\epsilon}{5\sqrt{3}}-\frac{\epsilon^{2}}{25})^\frac{1}{2} &\frac{\epsilon}{5}-\frac{1}{\sqrt{3}} &0 \\ 
\frac{1}{\sqrt{3}}-\frac{\epsilon}{5} & (\frac{2}{3}+\frac{2\epsilon}{5\sqrt{3}}-\frac{\epsilon^{2}}{25})^\frac{1}{2} & 0\\ 
0 & 0& 1
\end{pmatrix} 
\end{eqnarray}\\
We have,
\begin{align}
U_{PMNS}&=U(\theta_{23})U(\theta_{13})U(\theta_{12}) \nonumber \\
&= \begin{pmatrix}
u_{e1} & u_{e 2} & u_{e 3} \\ 
u_{\mu 1} & u_{\mu 2} & u_{\mu 3} \\ 
u_{\tau 1} & u_{\tau 2} & u_{\tau 3}
\end{pmatrix} 
\end{align}
where,
\begin{align*}
&u_{e1}=(1-\epsilon^{2})^\frac{1}{2} a(\epsilon),\\
&u_{e2}=(\frac{\epsilon}{5}-\frac{1}{\sqrt{3}})(1-\epsilon^{2})^\frac{1}{2},\\
&u_{e3}=\epsilon e^{-i \delta},\\
&u_{\mu 1}=\frac{1}{30}(5 \sqrt{3}-3 \epsilon)b(\epsilon)+\epsilon(\frac{1}{\sqrt{2}}-\frac{\epsilon}{2}) a(\epsilon) e^{i\delta},\\
&u_{\mu 2}=\frac{1}{30}\lbrace c(\epsilon) b(\epsilon)+\epsilon (\sqrt{2}-\epsilon)(3\epsilon - 5\sqrt{3}) e^{i\delta}\rbrace ,\\
&u_{\mu 3}=\frac{1}{2}(\epsilon - \sqrt{2})(1-\epsilon^{2})^{\frac{1}{2}},\\
&u_{\tau 1}=(\frac{1}{\sqrt{2}}-\frac{\epsilon}{2})(\frac{1}{\sqrt{3}}-\frac{\epsilon}{5})-\frac{1}{10}\epsilon b(\epsilon)d(\epsilon) e^{i \delta},\\
&u_{\tau 2}=(\frac{1}{\sqrt{2}}-\frac{\epsilon}{2})a(\epsilon)-\frac{1}{30}\epsilon(3 \epsilon - 5\sqrt{3})b(\epsilon) e^{i\delta},\\
&u_{\tau 3}= \frac{1}{2}(1-\epsilon^{2})b(\epsilon) ,
\end{align*}
and,
\begin{align*}
& a(\epsilon)=(\frac{1}{3}+\frac{2\epsilon}{5\sqrt{3}}-\frac{\epsilon^{2}}{25})^{\frac{1}{2}},\\
& b(\epsilon)=(2-\epsilon^{2} + 2\sqrt{2}\epsilon)^{\frac{1}{2}}, \\
& c(\epsilon)= (150+30\sqrt{\epsilon}-9 \epsilon^{2})^{\frac{1}{2}},\\
& d(\epsilon)=(\frac{50}{3}+\frac{10\epsilon}{\sqrt{3}}-\epsilon^{2})^{\frac{1}{2}} .
\end{align*}
It can be checked that,\\
\begin{eqnarray}
[U(\theta_{ij})]^{\dagger}U(\theta_{ij})=[U_{PMNS}]^{\dagger}U_{PMNS}=\begin{pmatrix}
1 &0 & 0 \\ 
0 & 1 &0 \\ 
0& 0 & 1 
\end{pmatrix} 
\end{eqnarray}
and we get,
\begin{eqnarray}
\tan^{2}\theta_{12}&=&\mid\frac{u_{e2}}{u_{e1}}\mid ^{2}=\frac{25-10\sqrt{3}\epsilon + 3\epsilon^{2}}{50+10\sqrt{3}\epsilon - 3\epsilon^{2}},\\
\tan^{2}\theta_{23}&=&\mid\frac{u_{\mu 3}}{u_{\tau 3}}\mid ^{2}=\frac{2-2\sqrt{2}\epsilon +\epsilon^{2}}{2+2\sqrt{2}\epsilon -\epsilon^{2}}. 
\end{eqnarray}
After interpreting the above two relations in terms of $\sin\theta_{13}$, we have,
\begin{eqnarray}
\tan^{2}\theta_{12}&=&\frac{1}{2}-\frac{1}{2}\sin \theta_{13}+\frac{1}{4}\sin^{2}\theta_{13}-\frac{2}{25}\sin^{3}\theta_{13} ,\\
\tan^{2}\theta_{23}&=&1-\frac{13}{5}\sin\theta_{13}+2\sin^{2}\theta_{13}-2\sin^{3}\theta_{13} . 
\end{eqnarray}
$U_{PMNS}$ for $\epsilon = 0$ and $\epsilon = 0.156 $ are shown below,
\begin{align*}
&U_{T.B.M}=\begin{pmatrix}
\sqrt{\frac{2}{3}} &-\sqrt{\frac{1}{3}}  & 0 \\ 
\sqrt{\frac{1}{6}} &\sqrt{\frac{1}{3}}  &  -\sqrt{\frac{1}{2}}\\ 
 \sqrt{\frac{1}{6}}& \sqrt{\frac{1}{3}}  & \sqrt{\frac{1}{2}}
\end{pmatrix},\\ \\
&U=\begin{pmatrix}
0.8274 & -0.5395 & 0.156 e^{-i\delta}  \\ 
0.4245 + 0.0822 e^{i \delta} & 0.6511 -0.0536 e^{i \delta} & -0.6214  \\ 
0.3435-0.1015 e^{i\delta} &0.5270 + 0.0662 e^{i\delta} & 0.7678
\end{pmatrix}.  
\end{align*}
This is clear from the above analysis that $\tan^{2}\theta_{12}=0.5$ and $\tan^{2}\theta_{23}=1$, if $\sin\theta_{13}=\epsilon=0$ (T.B.M mixing). At $\epsilon=0.155$ (N.H), $0.156$ (I.H) (the best-fit value of $\sin\theta_{13}$)[ 6 ], we get $\tan^{2}\theta_{12}=0.425$ and $\tan^{2}\theta_{23}=0.657,0.654$, which are very close to the best fit results [ 6 ]: $\tan^{2}\theta_{12}=0.443$ (N.H or I.H) and $\tan^{2}\theta_{23}=0.628$ (N.H) and $0.644$ (I.H). This is shown in Fig .1. where the variations of $\tan^{2}\theta_{12}$ and $\tan^{2}\theta_{23}$ are plotted against $\sin\theta_{13}$. 
\begin{figure}
\begin{center}
\includegraphics[scale=1]{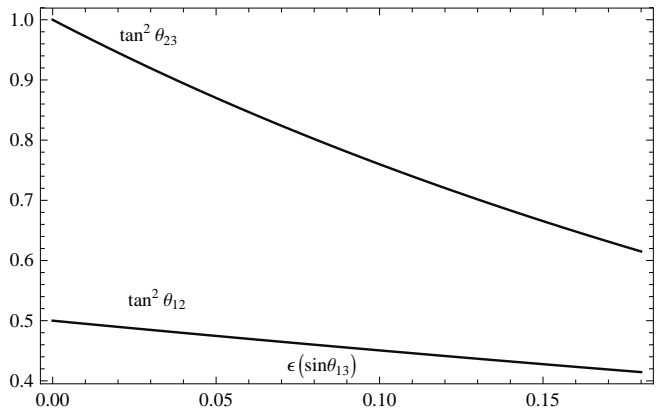} 
\caption{\footnotesize The variation of $\tan^{2}\theta_{12}$ and $\tan^{2}\theta_{23}$ with $\sin\theta_{13}$.}
\end{center}
\end{figure}
\section{Jarkslog parameter ( $J_{cp}$ )}
We introduce the CP phase $\delta$ in $U_{13}$ as shown in eq (8). The inclusion of $\delta_{cp}$ does not affect $\tan^{2}\theta_{12}$ or $\tan^{2}\theta_{23}$ [eq(9), eq(10)].
We obtain the $J_{cp}$ as,
\begin{align}
J_{cp} & = Im[u^{*}_{e1} u^{*}_{\mu 1} u_{e3} u_{\mu 1} ]\nonumber \\
\quad{}& = \epsilon (1-\epsilon^{2})(\frac{1}{\sqrt{2}}-\frac{\epsilon}{2})(\frac{1}{\sqrt{3}}-\frac{\epsilon}{5})(\frac{1}{2}+\frac{\epsilon}{\sqrt{2}}-\frac{\epsilon^{2}}{4})^{\frac{1}{2}} (\frac{2}{3}+\frac{2\epsilon}{5\sqrt{3}}-\frac{\epsilon^{2}}{25})^{\frac{1}{2}} \sin{\delta}
\end{align}
Maximum $J_{cp}$, i.e., $ J_{max}$ is obtained for $\delta = \frac{\pi}{2}$. For, $\epsilon = 0.156$, $J_{max}$ is obtained as 0.0341. The variation of $J_{max}$ with respect to $\sin\theta_{13}$ is shown in Fig.2 . Also the variation of $J_{cp}$ with $\delta$ ( with $\epsilon$ or $\sin\theta_{13}$ fixed at 0.156 ), is plotted in Fig.3. 
\begin{figure}
\begin{center}
\includegraphics[scale=1]{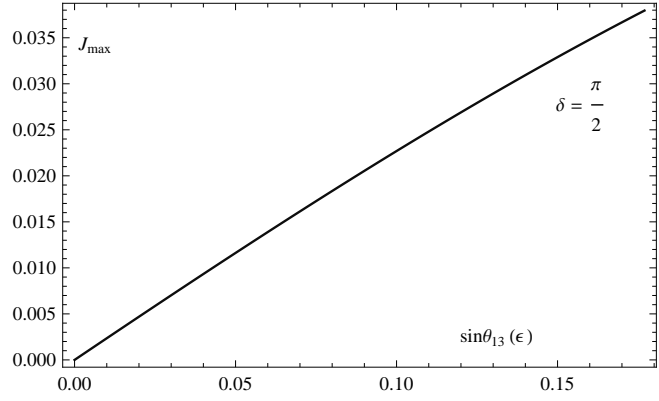} 
\caption{\footnotesize The variation of $J_{max}$ with $\sin\theta_{13}$.}
\end{center}
\end{figure}
\begin{figure}
\begin{center}
\includegraphics[scale=1]{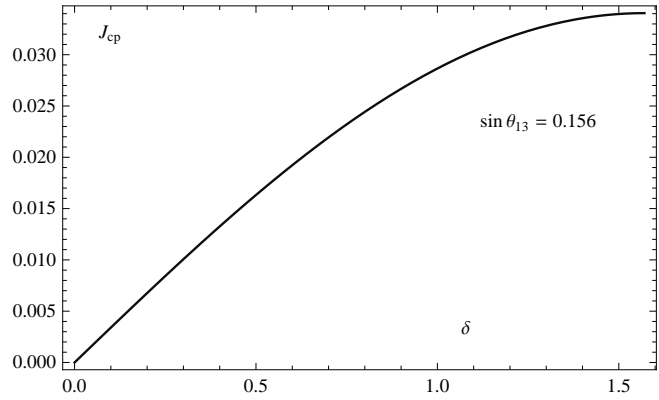} 
\caption{\footnotesize The variation of $J_{cp}$ with $\delta$. The range of $\delta$ is[ $0 \rightarrow \pi/2 $ ].}
\end{center}
\end{figure}
\section{Generation of neutrino mass matrix with inverted hierarchy}
We now apply the PMNS mixing matrix from eq (7), to construct the neutrino mass matrix with inverted hierarchy (I.H). We try to interpret the masses in terms of the same parameter $\sin\theta_{13}=\epsilon$.
We choose on phenomenological ground the absolute values of three neutrino masses in units of $eV$ as,
\begin{eqnarray}
m_{1}&=&\frac{60}{1250} + \frac{7}{8}\epsilon^{4},\\
m_{2}&=& \frac{61}{1250}+ \frac{7}{8}\epsilon^{4}-\frac{\epsilon^{5}}{3},\\
m_{3}&=&\frac{7}{8}\epsilon^{4}.
\end{eqnarray}
leading to,
\begin{eqnarray}
\Delta m^{2}_{21}&=& \frac{121}{1562500}+\frac{7\epsilon^{4}}{5000}-\frac{61}{1875}\epsilon^{5}-\frac{7}{12}\epsilon^{9}+\frac{1}{9}\epsilon^{10} ,\\
\Delta m^{2}_{23}&=& \frac{3721}{1562500}+\frac{427\epsilon^{4}}{5000}-\frac{61}{1875}\epsilon^{5}-\frac{7}{12}\epsilon^{9}+\frac{1}{9}\epsilon^{10} .
\end{eqnarray}
\begin{figure}
\begin{center}
\includegraphics[scale=1]{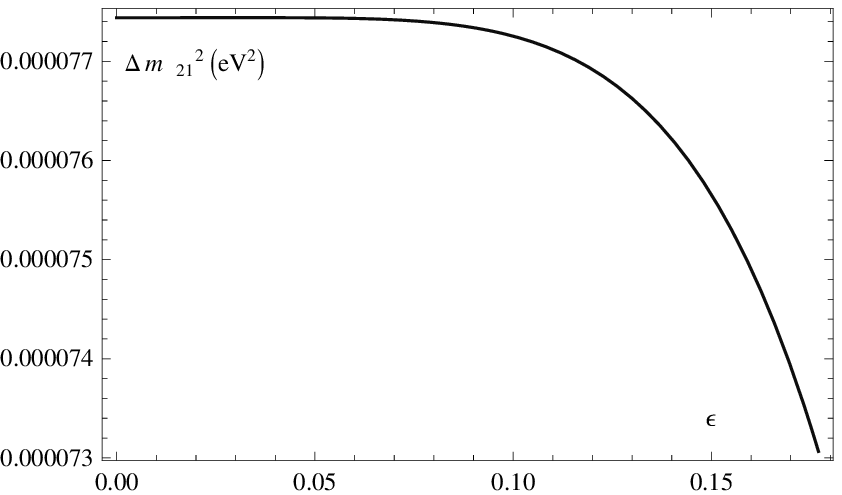} 
\caption{\footnotesize The variation of $\Delta m^{2}_{21}$ with $\epsilon (\sin\theta_{13})$.}
\end{center}
\end{figure}
\begin{figure}
\begin{center}
\includegraphics[scale=1]{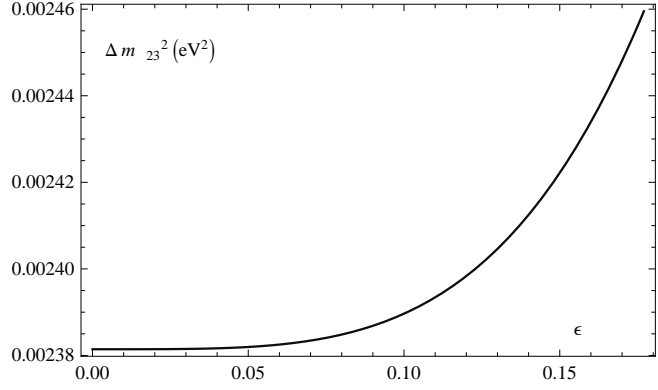} 
\caption{\footnotesize The variation of $\Delta m^{2}_{23}$ with $\epsilon (\sin\theta_{13})$.}
\end{center}
\end{figure}
At TBM mixing condition, i.e., at $\epsilon=0$, we get, $\Delta m^{2}_{21}=7.74\times10^{-5} eV^{2}$, $\Delta m^{2}_{23}=2.38\times10^{-3} eV^{2}$, and  $\Delta m^{2}_{21}=7.53\times10^{-5} eV^{2}$, $\Delta m^{2}_{23}=2.43\times10^{-3} eV^{2}$ are obtained at $\epsilon=0.156$.
For simplicity in the texture of the  neutrino mass matrix, we avoid the inclusion of $\delta_{cp}$. Using eqs ( 7 ) for $U_{PMNS}$  ( with $\delta_{cp}=0$ ) and eqs. 14 - 16 for $m_{i}$, we construct the neutrino mass matrix $M_{\nu}$ as follows,
\begin{align}
M_{\nu} &= U^{T}_{PMNS}.\begin{pmatrix}
m_{1} &0 &0 \\ 
0&-m_{2} & 0 \\ 
0&0 & m_{3}
\end{pmatrix}.U_{PMNS}=\begin{pmatrix}
m_{11} &m_{12} & m_{13} \\ 
m_{12} &m_{22} &m_{23} \\ 
m_{13} & m_{23} &m_{33} 
\end{pmatrix} 
\end{align}
where,
\begin{align*}
& m_{11}=\frac{7}{8}\epsilon^{6}+A^{2}(\epsilon)(1-\epsilon^{2})(\frac{6}{125}+\frac{7}{8}\epsilon^{4})-B(\epsilon)(1-\epsilon^{2})(\frac{1}{\sqrt{3}}-\frac{\epsilon}{5})^{2},\\
& m_{12}=-\frac{7}{8}\epsilon^{5}(1-\epsilon^{2})^{\frac{1}{2}}(\frac{1}{\sqrt{2}} -\frac{\epsilon}{2})+A(\epsilon)(1-\epsilon^{2})^{\frac{1}{2}}(\frac{6}{125}+\frac{7}{8}\epsilon^{4})\lbrace C(\epsilon)(\frac{1}{\sqrt{3}}-\frac{\epsilon}{5}) +\\
&\quad\quad\quad \epsilon A(\epsilon) (\frac{1}{\sqrt{2}}-\frac{\epsilon}{2})\rbrace+ B(\epsilon)(1-\epsilon^{2})^{\frac{1}{2}}(\frac{1}{\sqrt{3}}-\frac{\epsilon}{5})\lbrace \epsilon(\frac{\epsilon}{2}-\frac{1}{\sqrt{2}})(\frac{1}{\sqrt{3}}-\frac{\epsilon}{5})+C(\epsilon)A(\epsilon))\rbrace,\\
&m_{13}=\frac{7}{8}C(\epsilon)\epsilon^{5}(1-\epsilon^{2})^{\frac{1}{2}}+B(\epsilon)(1-\epsilon^{2})^{\frac{1}{2}}(\frac{1}{\sqrt{3}}-\frac{\epsilon}{5})\lbrace \epsilon C(\epsilon)(\frac{1}{\sqrt{3}}-\frac{\epsilon}{5})+A(\epsilon)(\frac{1}{\sqrt{2}}-\frac{\epsilon}{2})\rbrace \\
&\quad\quad\quad + A(\epsilon)(1-\epsilon^{2})^{\frac{1}{2}}(\frac{6}{125}+\frac{7\epsilon^{4}}{8})\lbrace(\frac{1}{\sqrt{2}}-\frac{\epsilon}{2})(\frac{1}{\sqrt{3}}-\frac{\epsilon}{5})-\epsilon A(\epsilon)C(\epsilon))\rbrace,\\
&m_{22}= \frac{7}{8}\epsilon^{4}(1-\epsilon^{2})(\frac{1}{\sqrt{2}}-\frac{\epsilon}{2})^{2} +(\frac{6}{125}+\frac{7\epsilon^{4}}{8})\lbrace C(\epsilon)(\frac{1}{\sqrt{3}}-\frac{2}{5})+\epsilon A(\epsilon)(\frac{1}{\sqrt{2}}-\frac{\epsilon}{2})\rbrace^{2}-\\
&\quad\quad\quad B(\epsilon)\lbrace \epsilon(\frac{\epsilon}{2}-\frac{1}{\sqrt{2}})(\frac{1}{\sqrt{3}}-\frac{\epsilon}{5})+A(\epsilon)C(\epsilon)\rbrace^{2},\\
& m_{23}=-\frac{7\epsilon^{4}}{8}C(\epsilon)(1-\epsilon^{2})(\frac{1}{\sqrt{2}}-\frac{\epsilon}{2})-B(\epsilon)\lbrace\epsilon C(\epsilon)(\frac{1}{\sqrt{3}}-\frac{\epsilon}{5})+A(\epsilon)(\frac{1}{\sqrt{2}}-\frac{\epsilon}{2})\rbrace \lbrace\epsilon(\frac{\epsilon}{2}-\\
&\quad\quad\quad\frac{1}{\sqrt{2}})(\frac{1}{\sqrt{3}}-\frac{\epsilon}{5})+C(\epsilon)A(\epsilon)\rbrace +\epsilon A(\epsilon)(\frac{6}{125}+\frac{7\epsilon^{4}}{8})\lbrace C(\epsilon)(\frac{1}{\sqrt{3}}-\frac{\epsilon}{5})+(\frac{1}{\sqrt{2}}-\frac{\epsilon}{2})\rbrace \\
&\quad\quad\quad\lbrace(\frac{1}{\sqrt{2}}-\frac{\epsilon}{2})(\frac{1}{\sqrt{3}}-\frac{\epsilon}{5})-\epsilon C(\epsilon)A(\epsilon)\rbrace,\\
&m_{33}=\frac{7\epsilon^{4}}{8} C(\epsilon)( 1-\epsilon^{2})-B(\epsilon)\lbrace \epsilon C(\epsilon)(\frac{1}{\sqrt{3}}-\frac{\epsilon}{5})+A(\epsilon)(\frac{1}{\sqrt{2}}-\frac{\epsilon}{2})\rbrace^{2}+(\frac{6}{125}+\frac{7 \epsilon^{4}}{8}) \lbrace (\frac{1}{\sqrt{2}}\\
&\quad\quad\quad -\frac{\epsilon}{2})(\frac{1}{\sqrt{3}}-\frac{\epsilon}{5})-\epsilon C(\epsilon)A(\epsilon)\rbrace^{2}.
\end{align*}
and,
\begin{align*}
&A(\epsilon)=(\frac{2}{3}+\frac{2\epsilon}{5\sqrt{3}}-\frac{\epsilon^{2}}{25})^{\frac{1}{2}},\\
&B(\epsilon)=(\frac{61}{1250}+\frac{7\epsilon^{4}}{8}-\frac{\epsilon^{5}}{3})^{\frac{1}{2}},\\
&C(\epsilon)=(\frac{1}{2}+\frac{\epsilon}{\sqrt{2}}-\frac{\epsilon^{2}}{4})^{\frac{1}{2}}.
\end{align*}
At, $\epsilon =0$ (T.B.M mixing), eq.( 22 ) reduces to $\mu - \tau $ symmetric mass matrix form ,
\begin{equation}
M_{\mu\tau} = \begin{pmatrix}
\delta_{1} & 1  & 1 \\ 
1 & \delta_{2}  & \delta_{2}  \\ 
1 & \delta_{2} & \delta_{2} 
\end{pmatrix} m_{0},
\end{equation}
with $\delta_{1,2} \ll 1$, for inverted hierarchy. From eq.(19) we
have the neutrino mass matrix and its mass eigenvalues,
\begin{eqnarray}
M_{\nu} & =& M_{\mu\tau}=\frac{121}{3750}\begin{pmatrix}
\frac{1}{2} & 1 & 1 \\ 
1 & -\frac{1}{4} & -\frac{1}{4} \\ 
1 & -\frac{1}{4} & -\frac{1}{4}
\end{pmatrix} =\begin{pmatrix}
0.0157& 0.0323 & 0.0323 \\ 
0.0323 & -0.0083& -0.0083 \\ 
0.0323 & -0.0083 & -0.0083\\
\end{pmatrix} , \nonumber\\
m_{1}&=&\frac{60}{1250} ,\quad m_{2}=-\frac{61}{1250}, \quad m_{3}= 0 \quad \text{in $eV$}.
\end{eqnarray}

For  $\epsilon=0.156$, eq.(19) leads to
\begin{align}
M_{\nu}&= M_{\mu\tau} + \Delta M_{\nu} \nonumber\\
\quad {} &=\begin{pmatrix}
 0.0189 & 0.0362 & 0.0255 \\ 
 0.0362 & -0.0049 & -0.0118 \\ 
 0.0255& -0.0118 & -0.0142
\end{pmatrix}
\end{align}
where,
\begin{align*}
&\Delta M_{\nu} =\begin{pmatrix}
 0.0032 & 0.0039  & -0.0068 \\ 
 0.0039 & 0.0034 & -0.0035 \\ 
-0.0068 & -0.0035  & -0.0059
\end{pmatrix}\\
&m_{1}=0.0485 ,\quad m_{2}=-0.0493 ,\quad m_{3}= 0.0005 \quad \text{in $eV$}.
\end{align*}
\section{Summary}
We have started with a parameter $\epsilon$ equating this to $\sin\theta_{13}$ and construct the PMNS matrix, $U_{PMNS}$. Then we represent the neutrino masses ($m_{i=1,2,3}$) in terms of the same parameter $\sin\theta_{13}$, i.e $\epsilon$. We verify our hopothesis by comparing the ranges of the  mass squared differences as a result of our ansatz  with the  $1\sigma$ range, experimentally obtained. We take the range of $\epsilon$ as the experimental $1\sigma$ range of $\sin\theta_{13}$ [6]. We obtain the range of $\Delta m^{2}_{21}$ and $\Delta m^{2}_{23}$ as $( 7.46-7.58 )\times 10^{-5} eV^{2}$ and $( 2.42-2.44 ) \times 10^{-3} eV^{2}$ respectively. The respective ranges obtained, lie within the experimental $1\sigma$ boundary [6]. This provides a support to our hypothesis $m_{i}$ as $m_{i}(\epsilon)$. This is to be emphasised that the $U_{PMNS}$ matrix as proposed in eq.(7) satisfy the unitary condition and is not dependent on the choice of the order of $\epsilon$. The introduction of $\delta_{cp}$ does not affect $\tan^{2}\theta_{12}$ and $\tan^{2}\theta_{23}$ in our calculation. The maximum $J_{cp}$ obtained is 0.034 ( with respect to $\epsilon = \sin\theta_{13} = 0.156$ ). Finally we concentrate on the construction of $M_{\nu}$, the neutrino mass matrix. The present investigation though phenomenological, gives a complete picture of the texture of the neutrino mass matrix which can be employed in other applications regarding  baryon asymmetry of the Universe [18]. Although we have constructed the mass matrix for inverted hierarchical model, yet we can extend our technique to Normal as well as Quasidegenerate mass models.

\end{document}